\documentclass[prb, 10pt, twocolumn]{revtex4}

\usepackage[english]{babel}
\usepackage[intlimits]{amsmath}
\usepackage{amssymb}
\usepackage{amsthm}
\usepackage[dvips]{graphicx}
\usepackage{setspace}

\begin{document}

\begin{titlepage}

\begin{spacing}{1.0}

\title{Plasticization and antiplasticization of polymer melts diluted by low molar mass species}

\author{Evgeny B. Stukalin$^1$, Jack F. Douglas$^2$, and Karl F. Freed$^1$}

\affiliation{$^1$The James Franck Institute and Department of Chemistry, University of Chicago, 
Chicago, Illinois 60637 \\
$^2$Polymers Division, National Institute of Standards and Technology, Gaithersburg, Maryland 20899}

\begin{abstract}

\bigskip

An analysis of glass formation for polymer melts that are diluted by structured molecular additives is derived by using the generalized entropy theory, which involves a combination of the Adam-Gibbs model and the direct computation of the configurational entropy based on a lattice model of polymer melts that includes monomer structural effects. Our computations indicate that the plasticization and antiplasticization of polymer melts depend on the molecular properties of the additive. Antiplasticization is accompanied by a ``toughening'' of the glass mixture relative to the pure polymer, and this effect is found to occur when the diluents are small species with \textit{strongly attractive} interactions with the polymer matrix. Plasticization leads to a decreased glass transition temperature $T_g$ and a ``softening'' of the fragile host polymer in the glass state. Plasticization is prompted by small additives with \textit{weakly attractive} interactions with the polymer matrix. The latter situation can lead to phase separation if the attractive interactions are sufficiently strong. The shifts in $T_g$ of polystyrene diluted by fully flexible short oligomers (up to 20 $\%$ mass of diluent) are evaluated from the computations, along with the relative changes in the isothermal compressibility at $T_g$ (a ``softening'' or ``toughing'' effect) to characterize the extent to which the additives act as antiplasticizers or plasticizers.  The theory predicts that a \textit{decreased} fragility can accompany both antiplasticization and plasticization of the glass by molecular additives. The general reduction in the $T_g$ and fragility of polymers by these molecular additives is rationalized by analyzing the influence of the diluent's properties (cohesive energy, chain length, and stiffness) on glass formation in diluted polymer melts. The description of glass formation at fixed temperature that is induced upon increasing the diluent concentration directly implies the Angell equation ($\tau_{\alpha} \sim A \, \text{exp}\{ B/(\phi_{0,\, p} - \phi_p) \}$) for the structural relaxation time as function of the polymer concentration, and the computed ``zero mobility concentration'' $\phi_{0, \, p}$ scales linearly with the inverse polymerization index $N$.

\end{abstract}

\maketitle

\end{spacing}

\end{titlepage}

\begin{spacing}{1.0}

\section{Introduction}	

Recent computations using the generalized entropy theory of glass-formation indicate that the strong temperature dependence of the structural relaxation in polymer glass-forming liquids is due to packing frustration associated with complex monomer shapes, chain connectivity, and the stiffness of the side groups and/or backbone. \cite{dudowicz08} Weitz and coworkers have likewise established that packing efficiency (which is modified by changing rigidity) governs fragility in model soft colloidal particle systems, \cite{mattsson09} so that packing efficiency in the glass state seems to be generally implicated in affecting fragility. The entropy theory of glass formation (ETGF) provides an understanding of many other observed trends in the variation of the fragility of polymer glass former with chain stiffness, cohesive energy, and chain length. \cite{stukalin09} Moreover, the calculations reproduce the ``internal plasticization" \cite{jackson67, merrill65} effect in poly($\alpha$-olefins) that appears as the length of flexible side groups increases. \cite{stukalin09} Our computations also provide solid support for the interpretation and empirical generalization of experimental findings concerning the influence of molar mass and pressure on glass formation. The theory further implies that profound changes in the properties of glass-forming fluids can be induced by modifying chain packing through film confinement or the addition of diluents, and the theory provides a qualitative framework for understanding these changes in terms of variations in molecular packing. 

Usually a diluent, i.e., a ``plasticizer", significantly decreases the glass transition temperature $T_g$ from that of a bulk polymer. Early measurements indicate that small molecule diluents (solvents) with a lower $T_g$ depress the $T_g$ of the host polymer to a greater extent than solvents with higher $T_g$. \cite{jenckel53, plazek96} Molecular plasticizers often impart increased flexibility to the polymer in the glass state while also reducing the glass transition temperature. In contrast to the more widely studied plasticizers, antiplasticizers also depress $T_g$, but these additives \textit{increase} the stiffness (shear or bulk modulus) of polymeric materials in the glass state. \cite{maeda87b, vrentas88, cauley91, ngai91} This stiffening is of great importance for preservation of biomaterials, such as foods, tissues, and drugs, and for enhancing the scratch resistance of polymer films, controlling the brittleness and other non-linear mechanical properties of polymer materials, etc. Thus, antiplastization has many practical applications. \cite{caliskan04, moznine03} Examples of antiplasticizer/polymer pairs involve tricresyl phosphate in polysulfonate \cite{maeda87a} and dibutylphthalate in polycarbonate. \cite{casalini00}

Using a lattice model of polymer glass-formation in which an ideal glass transition is identified with the extrapolated vanishing of the configurational entropy, DiMarzio and coworkers \cite{dimarzio63, dimarzio96} conclude that the main effect of a solvent additive is to depress the glass transition temperature. In particular, they find that the depression of $T_g$ is a nearly a universal linear function of the plasticizer \textit{mole fraction}. Recent molecular dynamics simulations have examined the relaxation of bulk polymer-solvent mixtures \cite{peter09a} and supported polymer films diluted by a good solvent \cite{peter09b} near the glass transition temperature. The calculations also indicate that $T_g$ decreases linearly with diluent concentration (volume fraction) over a wide range of solvent concentrations (up to 25 $\%$). \cite{peter09a} These earlier theoretical and computational works, however, do not explore the influence of diluent structure and thermodynamic interactions with the polymer, and they also do not describe changes in the fragility of glass formation. 

The present work considers oligomeric molecular additives and thus is not restricted to ``structureless'' small molecule diluents. We also study systems with a wide range of polymer-diluent interaction energies to capture general physical trends in polymer/diluent glass-formation that are associated with molecular size and structure. Our study thus requires the computationally non-trivial extension of the generalized entropy theory (ETGF) to describe these complex mixtures and thereby reveal molecular characteristics that promote plasticization and antiplasticization. The analysis is limited to miscible fluid mixtures having a moderate concentration of additives (about 10 $\%$ mass) where only a single calorimetric glass transition exists. The compositional variation of $T_g$ (and fragility) is evaluated for this relative low range of concentrations for miscible binary fluids.
 
The generalized entropy theory of Dudowicz, Douglas, and Freed (DDF) \cite{freed03, dudowicz06, dudowicz08} focuses mainly on the temperature range above $T_g$, where a thermodynamic description of glass-formation is theoretically sensible. The entropy theory is generalized here to describe glass formation in multi-component fluids (e.g., binary mixtures \cite{freed03}), provided a single glass transition appears in the multi-component fluid. 

The ETGF describes the influence on thermodynamic properties of short-range correlations induced by chain connectivity, semi-flexibility, and monomer structure. The more realistic description of the thermodynamics of polymeric systems is achieved using the lattice cluster theory (LCT) generalization of the Flory theory for semi-flexible chains. This ETGF is combined with Adams-Gibbs theory \cite{adam65, mohanty94} to describe the structural relaxation times,
\begin{equation}\label{tauintro}
\tau_{\alpha} = \tau_0 \, \text{exp} \{ \beta \Delta \mu [s_c^{*}/s_c(T)] \},
\end{equation}
where $\beta = 1/k_B T$,  $\tau_0$ is the high temperature limit of the relaxation time, $\Delta \mu$ is the activation energy at high temperatures, and $s_c^{*}$ is the high temperature limit (actually peak value) of the configurational entropy density $s_c(T)$. Over the temperature range $T_g < T < T_I$, the configurational entropy density is found to obey the relation $s_c (T) = s_0 (1 - T_0/T)$ that implies, 
$ \tau_{\alpha} = \tau_0  \, \text{exp} \{ \Delta \mu \, s_c^{*}/[s_0 \, k_B (T - T_0)] \}$, where the  fragility parameter at low temperatures is determined from the relation, 
\begin{equation}
D = [ \Delta \mu /(k_B T_0)][s_c^{*}/s_0].
\end{equation}
DDF use the empirically based relation between $\Delta \mu$ and $T_I$, namely $\Delta \mu/k_B \simeq 6 T_I$, to compute the fragility parameter $D$ without adjustable parameters beyond those used in the LCT for the thermodynamics. \cite{dudowicz05b} (This procedure greatly simplifies the description of mixtures since otherwise the composition dependence of $\Delta \mu$ would have to be established on a system by system basis.) The characteristic temperature $T_I$ is determined as the inflection point in $T s_c(T)$, marking the crossover from the range $T_g < T < T_I$, where the product $T s_c(T)$ varies linearly with $T$ \cite{dudowicz05a, dudowicz05b} to the region of higher temperatures $T > T_I$ at which the behavior is clearly non-linear. Experimental \cite{voronel97, veliyulin95, bottezzati89, hayashiuchi82} and simulation \cite{sastry98, riggleman09} data for various glass-formers suggest that $\Delta \mu/k_B$ is approximately six times the experimental mode coupling temperature $T_{mc}$, which is thus  identified in our studies with $T_I$. The reported high temperature activation energies $\Delta \mu$ for segmental relaxation of relatively fragile polymers vary from 12.1 kJ/mol (polybutylene, $m = 85$) to 17.2 kJ/mol (poly(methyl acrylate), $m = 102$) and tend to be higher for strong polymers. \cite{sokolov07}

The first calculation of the glass transition temperature by DDF \cite{dudowicz05a} uses a Lindemann criterion \cite{xia00, lindemann10, bilgram87, lowen94, laviolette85} defined in terms of a critical mean square displacement relative to the average interparticle separations. However, the resultant $T_g$ are quite close to those obtained from the conventional (and less controversial) definition of $T_g$ as the temperature at which the computed structural relaxation time equals a value characteristic of glass formation \cite{dudowicz08}, e.g., 100 s. Hence, the present work uses the latter simpler and common definition.

Section II focuses on implementing the ETGF to describe vitrification, using polystyrene (PS) melts as a typical example of a relatively fragile polymer for which the influence of different solvents on glass formation is well documented. The entropy theory for binary mixtures is then applied to examine how the properties, such as the size and interaction energies, of diluents added to a fragile polymer affect the nature of glass-formation (section III). Section IV compares the general trends in glass formation for binary fluids and polymer melts. The ETGF is then applied in Sect. V to estimate the efficiency parameters for (anti-) plasticizers, where PS is used as the host polymer. Section VI describes the compositional variation of the structural relaxation times and glass transition temperatures of the plasticized polymer melts.  The temperature and chain length dependence of the extrapolated ``zero mobility'' polymer concentration (the analog of the Vogel-Fulcher-Tammann (VFT) temperature $T_0$ in glass formation upon cooling) is discussed briefly in section VII.

\section{Application of entropy theory to species with different rigidities of backbone and side groups: polystyrene as a typical example} 

Because cyclic portions of monomer structures have not yet been implemented in the entropy theory of glass formation, a ``compact'' structure is introduced to mimic the size and shape of a polystyrene (PS) monomer whose structure is taken here as comprising two backbone carbons and six side group carbons (see Fig. 1) (each with their bonded hydrogen atoms as united atom groups occupying single lattice sites). 
\begin{figure}[ht]
\renewcommand{\figurename}{FIG.}
\begin{center}
\unitlength 1in
\includegraphics[width=0.800in]{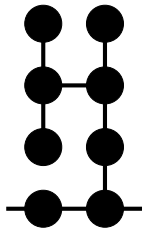}
\caption{\label{fig1} Representation of the styrene monomer as a collection of united atom groups (CH$_n$) used in the lattice cluster theory.}
\end{center}
\end{figure}
One bending energy $E_b$ describes the local flexibilities of the backbone, and another $E_s$ describes the side group flexibility.  A single effective van der Waals energy parameter $\epsilon$ is used for all carbon atoms to minimize the number of adjustable parameters, although the actual energies vary.  Hence, the cohesive energy $\epsilon$ and the cell volume $a_{cell}$ are the two quantities describing intermolecular polymer interactions and determining the equation of state (EOS) of a polymer melt in the entropy theory. 
The bending energies only exert a minor influence on the specific volume, thermal expansion coefficient, and isothermal compressibility. Thus, given the particular molecular topology in Fig. 1, the parameters $\epsilon$ and $a_{cell}$ are fit to the specific volumes of PS in the range from 350 K to 450 K. Our interest focuses on experimental measurements in a $\Delta T = 100$ K region above $T_g$. \cite{orwoll96, zoller82} The fitting yields the cohesive energy $\epsilon = 275$ K and cell volume $a_{cell} = 2.72$ $\AA$ parameters for high molar mass PS melts ($N = 8000$). The backbone and side group bending energies of high molar mass PS are chosen to reproduce the experimental $T_g = 375$ K and fragility $m = 139$. \cite{plazek91, bohmer93, qin06} (Experiments have $T_g$ varying from 360 K to 375 K, and $m$ is reported by different authors as $m = 116$ \cite{huang01}, 124 \cite{schwartz07}, 143 \cite{roland99}, 146. \cite{santangelo98} The reported or extracted VFT parameter $D = 1/K$ is more scattered, as D = 3.83 \cite{schwartz07}, $4.75$ \cite{hintermeyer08}, 5.77 \cite{sahnoune96}, and 7.80. \cite{santangelo98}) The determination of $E_b$ and $E_s$ from experimental data is facilitated by the observation that $m$ and $T_g$ are linearly correlated for a given ratio $E_b/E_s$ and also that the side groups in PS are stiffer than the backbone. The resultant bending energies are determined as $E_b = 650$ K and $E_s = 1300$ K, giving the ratio $E_s/E_b \simeq 2$.

\section{Influence of diluent's properties on glass formation in binary mixtures}

The influence of different additives on glass formation in fragile high molar mass polymers is examined by first considering the SF (``stiff'' backbone, ``flexible'' side groups) type polymer, ($N = 8000$, $E_b = 900$ K, $E_s = 450$ K) diluted by a relatively small amount (10 $\%$ mass) of a fully flexible ($E_b = E_s = 0$ K) oligomeric species with variable properties. The calculations here take the monomer in both the SF polymer and the additive to have the 1-pentene structure (two united atom groups in the backbone and three united atom groups in the side group \textit{per} monomer). The computed glass transition temperature and fragility of the undiluted SF polymer are $T_{g, \, p} = 321$ K and $K_p = 1/D_p = 0.30$. Alternatively, the fragility of the SF melt is calculated as $m_p = 194$ from the relation $m = A + \text{ln} \,10 \, A^2/D$ where $A = \text{log} \, \tau_{\alpha}(T_g) - \text{log} \, \tau_0$. For computational purposes, $\tau_0 = 10^{-14}$ s is used as a typical value for the relaxation time limit at high temperatures, and the $T_g$ of both melts and binary fluids is determined conventionally as the temperature for which the computed structural relaxation time is $\tau_{\alpha} = 100$ s.

\subsection{Influence of the size of the additive}
The depression in the $T_g$ of the mixture relative to the undiluted polymer melt is found to increase as the size of the additive diminishes and to saturate as the size of the diluent becomes large enough. 
\begin{figure}[ht]
\renewcommand{\figurename}{FIG.}
\unitlength 1in
\begin{center}
\includegraphics[width=3.0in]{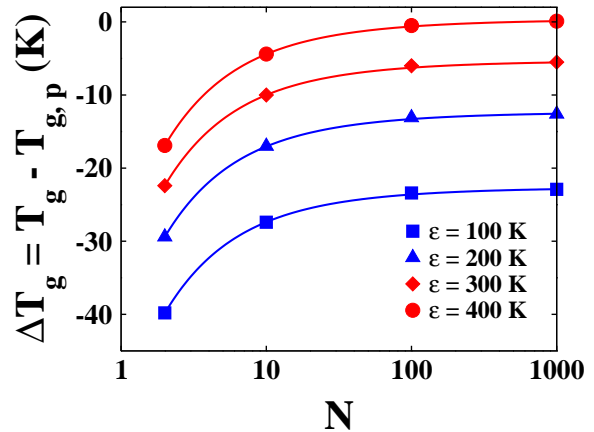}
\caption{\label{fig2} Shifts in $T_g$ for a SF polymer melt diluted by 10 $\%$ mass of oligomeric additives as a function their polymerization index $N$. Each curve corresponds to a specific cohesive energy $\epsilon$ of the diluents.}
\end{center}
\end{figure}
Thus, small oligomeric species (monomers, dimmers) exert the most pronounced depression of $T_g$. Figure 2 depicts the dependence of $T_g$ of the semi-flexible SF polymer diluted by 10 $\%$ mass of fully flexible oligomers with variable polymerization indices $N$.
\begin{figure}[ht]
\renewcommand{\figurename}{FIG.}
\begin{center}
\unitlength 1in
\includegraphics[width=3.0in]{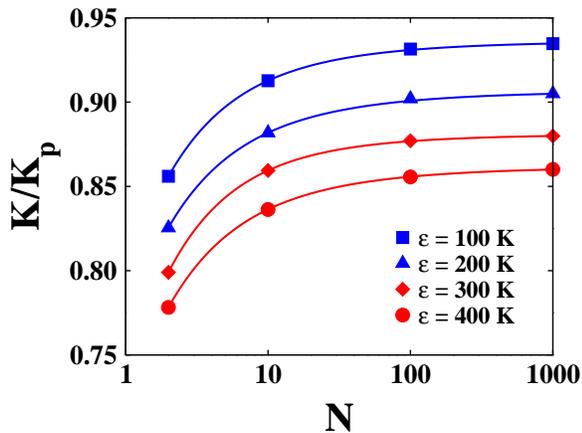}
\caption{\label{fig3} Influence of size and cohesive energy of the additives on the fragility of a SF polymer melt mixed with 10 $\%$ mass of the oligomeric diluent.}
\end{center}
\end{figure}
The fragility of these mixtures also decreases, as depicted in Fig. 3. The diminution of the fragility $K$ as the additive ``shrinks'' in size (and $N$ decreases) is accompanied by a decrease in the computed fraction of excess free volume $\phi_v$ at $T_g$, a relation previously quantified by us for polymer melts with variable molar mass $M$. The isothermal compressibility $\kappa$ changes insignificantly as $N$ varies, unlike $T_g$ (see Fig. 4), and similarly, the specific density $\rho$ of the mixture has no appreciable dependence on the sizes of the oligomeric additives (see Fig. 5)
\begin{figure}[ht]
\renewcommand{\figurename}{FIG.}
\unitlength 1in
\includegraphics[width=3.0in]{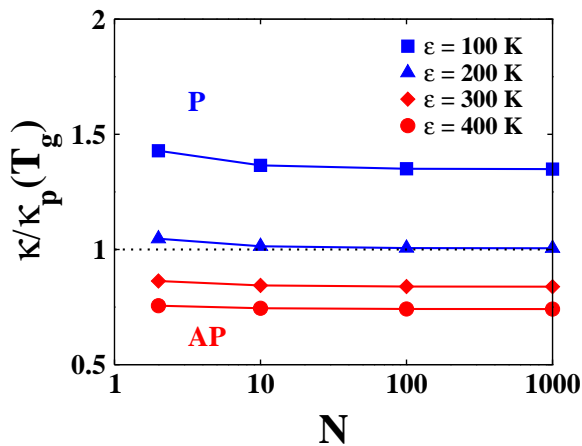}
\caption{\label{fig4} Ratios of the isothermal compressibility of the mixture (SF host polymer + 10 $\%$ mass of the additive) to the pure SF melt evaluated at the $T_g$ of the mixture as a function of size $N$ of the oligomeric additives. Each curve corresponds to the specified cohesive energy $\epsilon$ of the diluents.}
\end{figure}
\begin{figure}[ht]
\renewcommand{\figurename}{FIG.}
\begin{center}
\unitlength 1in
\includegraphics[width=3.0in]{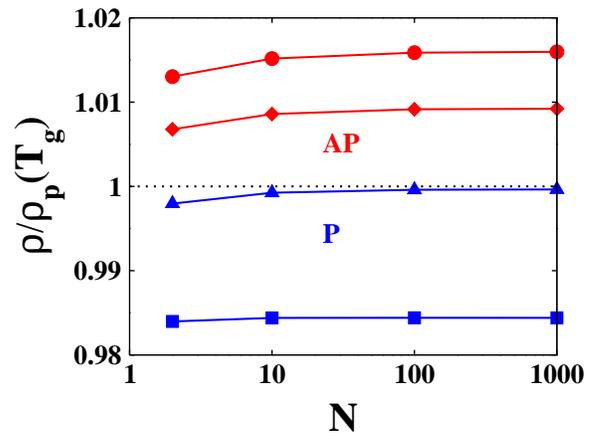}
\caption{\label{fig5} Ratios of specific density of the mixture (SF host polymer + 10 $\%$ mass of the additive) and the pure SF melt at the $T_g$ of the mixture as a function of the size $N$ of the oligomeric additives. Each curve corresponds to the specific cohesive energy $\epsilon$ of the diluents.}
\end{center}
\end{figure}
Thus, decreasing the size of the (fully flexible) diluent produces a larger depression of $T_g$ and renders the mixture a stronger glass-former than the undiluted polymer. However, varying the diluent's size and flexibility has only a small influence on the ``softness'' of the material, as measured by its isothermal compressibility $\kappa$ and specific density $\rho$. The change in the stiffness of the antiplasticizer or plasticizer additive is then predicted to arise mainly from a change in the average cohesive energy density by the additive.

\subsection{Influence of cohesive energy of the diluent}

The polymer-diluent van der Waals interactions $\epsilon_{ps}$ are taken as the usual geometric average of the pure component energies, $\epsilon_{ps} = (\epsilon_{pp} \epsilon_{ss})^{1/2}$. The computations fix the van der Waals interaction strength of the host polymer at $\epsilon_{pp} = 200$ K \cite{foreman97}, while the diluent energy $\epsilon_{ss} = \epsilon$ is varied. Figures 2 and 3 display the dependence of $T_g$ and the fragility $K$, respectively, of mixtures comprising the semi-flexible SF polymer diluted by 10 $\%$ mass of the fully flexible additives with different cohesive energies $\epsilon$. Figure 4 presents the analogous dependence of the ratio $\kappa/\kappa_p$ of the isothermal compressibilities of the mixture and the melt evaluated at $T_g$ (see the caption).  The depression of $T_g$ for the mixture relative to the undiluted host polymer diminishes as the cohesive energy of the additive's $\epsilon$ increases (at fixed size of the diluent $N$), while the fragility $K$ of the mixture drops relative to the melt. Thus, increasing the additive's cohesive energy $\epsilon$ (at fixed size $N$) produces a stronger glass-former compared to the undiluted melt but diminishes the depression of $T_g$.  The shift $\Delta T_g$ is approximately linear in $\epsilon$ at fixed $N$, analogous to the variation of $T_g$ with $\epsilon$ for polymer melts found by us previously. \cite{stukalin09} 

Recent work by Riggleman \textit{et al.} \cite{riggleman09} indicates that the addition of antiplasticizing molecular additives to polymer materials reduces the compressibility of the mixture and correspondingly increases the shear and bulk moduli in the glass state, while glass plasticizing additives can be defined as molecular additives that have the opposite effect. It is important to realize, however, that antiplasticization normally exerts an opposing influence on the shear modulus (high frequency modulus, of course) in the melt state, so that the best antiplasticizers of the mixture in the glass state are also the best plasticizers in the fluid domain. \cite{kinjo73} Thus, discussions of plasticization and antiplasticization must also specify the temperature regime considered. \cite{starr09b, psurek08} The present paper strictly refers to plasticization and antiplasticization of the glass state, and we use the compressibility and density \textit{together} as suitable measures of this effect. The computed compressibility and density do not exhibit an inversion in relative values above and below $T_g$, as observed for the shear modulus in some systems, \cite{maeda87c, starr09b, psurek08, cukierman91} and thus cannot provide reliable information about material ``softness'' in the melt state. Riggleman \textit{et al.} \cite{riggleman09} have not yet studied the case of small molecule additives with weak attractive interactions to explore the properties of melts with plasticizing additives.

The shift in the isothermal compressibility (density variance) of the diluted polymer depends on the intermolecular interactions in the binary fluid: as the cohesive energy $\epsilon = \epsilon_{ss}$ of the additive grows (and, hence, $\epsilon_{ps}$ for heterocontact interactions also increases), the mixture's compressibility diminishes. More specifically, our model computations indicate that when $\epsilon = \epsilon_{ss} > \epsilon_{pp}$, the compressibility of the mixture diminishes with the additive (as does the ``softness'' of the material), while the opposite limit of $\epsilon = \epsilon_{ss} < \epsilon_{pp}$ leads to an increasing relative compressibility (the ``softness'' becomes larger on dilution). [The decrease of the compressibility (variance of density fluctuations) has been observed in connection with the suppression of the $\beta$-relaxation caused by molecular additives \cite{fischer85}, an effect suggested to be related to the antiplasticization phenomenon; see discussion in ref. by Bergquist \textit{et al.} \cite{bergquist99}] The influence of $\epsilon$ on the relative ``softness'' of the mixtures enables us to classify flexible diluents that $\it{depress}$ $T_g$ into two typical classes: glass plasticizers and antiplasticizers. If the flexible diluent species are relatively small (low $N$), the depression of $T_g$ is greatest (at fixed $\epsilon$). The addition of toluene to PS is probably a good example: \cite{ngai91b} $T_g$ drops, and the system becomes stronger with polymer dilution. When $\epsilon$ exceeds $\epsilon_{pp}$ of the host polymer, the material becomes ``toughened'' in the glass state, a characteristic feature of glass antiplasticization, while decreasing $\epsilon$ leads to ``softening'' of the mixture, a characteristic property of glass plasticization. These changes in compressibility as $\epsilon$ varies appear for all temperatures.  The specific density $\rho$ of the mixture (evaluated at $T_g$) modestly increases as an antiplasticizer (large $\epsilon$) is added to the melt, while $\rho$ tends to diminish when a plasticizer (low $\epsilon$) is mixed with the polymer (see Fig. 5). This effect, which has often been observed, is not by itself sufficient to predict whether an additive is an antiplasticizer or not. \cite{maeda87b, maeda87a} Additional information is required concerning the density and density variance. As $\epsilon$ increases (at fixed $N$), the fragility $K$ of the mixture decreases (Fig. 4) in parallel with the diminishing excess free volume fraction $\phi_v$ at $T_g$ in a similar fashion as found previously for polymer melts. \cite{stukalin09}

We conclude that the ``best'' glass plasticizers are small flexible species with \textit{weak attractive} interactions, while the ``best'' antiplasticizers are also compact, but additionally have \textit{strongly attractive} interactions with the polymer matrix (polar interactions are often especially effective, but the presence of halogen, nitrogen, oxygen and sulfur groups can also be effective in creating these relatively strong interactions), as noted from previous experimental studies. \cite{jackson67, jackson67b} Increasing $\epsilon$ ``toughens'' the material but diminishes the depression of $T_g$, so additives with very attractive interactions are expected to produce a small depression in $T_g$ (or even an increase of $T_g$). Hence, the size $N$ and the flexibility (as described in the next section) of the additives play primary roles in governing a mixture's $T_g$, but have little impact on material ``softness'' (as determined by the relative compressibility). The cohesive energy $\epsilon$ of the diluent changes both $T_g$ and the ``softness'' of a material. Dong and Fried \cite{fried97} utilize a statistical thermodynamic theory to describe the $T_g$s of polymers \textit{plasticized} by different molecules and arrive at similar conclusions, i.e., high efficiency plasticizers tend to exhibit a weak interactions with the host polymer and are small in size relative to the polymer statistical segment size.

\subsection{Influence of semi-flexibility of the additives on glass formation}

As the stiffness of the additive molecules increases, the depression of $T_g$ of the mixture diminishes progressively and can even produce an increase in $T_g$. We have computed the ideal glass transition temperature $T_0$, where configurational entropy extrapolates to zero, for mixtures of semi-flexible SF polymers and 10 $\%$ mass of oligomeric additives for a range of diluent's sizes ($N = 2$, 10), cohesive energies ($\epsilon = 100$ K, 400 K), and a wide range of bending energies $E_b$ (0 K to 2000 K). The ratio of the backbone and side group bending energies is kept fixed at the same ratio as in the host semi-flexible polymer, i.e., $E_b/E_s = 2$. $T_0$ varies with $E_b$ in a sigmoidal form, leveling off at low and high diluent rigidities (see Fig. 6).
\begin{figure}[ht]
\renewcommand{\figurename}{FIG.}
\begin{center}
\unitlength 1in
\includegraphics[width=3.0in]{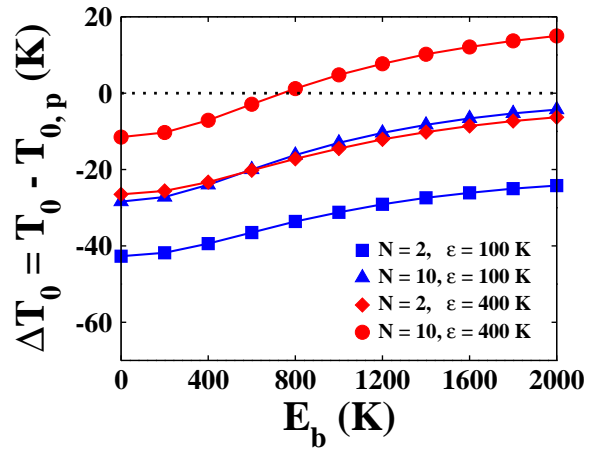}
\caption{\label{fig6} Shifts in the ideal glass transition temperatures $T_0$ for a SF polymer melt diluted by oligomeric additives (10 $\%$ mass) with different sizes $N$ and cohesive energies $\epsilon$ as a function of the bending energy $E_b$ of the diluent backbone. The stiffness of the side groups in the additives is $E_s = \text{1/2} E_b$ (see text).}
\end{center}
\end{figure}
Evidently, the influence of the local stiffness of the additive on $T_0$ (and $T_g$) of the mixture is greater for diluents with higher molar masses since the diluents having larger monomers contain a greater number of stiff bond pairs. (Computations for fully flexible additives display a linear correlation between $T_0$ and $T_g$. The strength of this correlation supports the assumption that the trends in the variation of $T_0$ as a function of stiffness also apply to $T_g$).

\section{Relation between the general trends in glass formation for binary fluids and polymer melts: scaling relation}

The computed variation of $\Delta T_g$ for mixtures with $\epsilon$ and $N$ (see Fig. 2) can be scaled into the general relation,
\begin{equation}\label{fitepsn}
\Delta T_g (N, \epsilon) = - \phi_s T_{g, \, p} \, \text{ln} \left (  \frac{a + b/N}{1 + k \, \epsilon}  \right),
\end{equation}
where $a$, $b$ and $k$ are coefficients that are independent of $\epsilon$ and $N$. The general scaling relation for $\Delta T_g$ of mixtures as a function of $N$ and $\epsilon$ may be rationalized based on the dependence of $T_g$ for an (oligomeric) diluent on its polymerization index $N$ and the strength of the interaction potential $\epsilon$. The form of this relation is ``derived'' in section VI which focuses on the concentration dependence of $T_g$ for the mixtures.  Equation (\ref{fitepsn}) successfully reproduces the depression of $T_g$ when the host high molar mass polymer is diluted by different oligomeric fully flexible additives. (The caption of Fig. 7 presents the numerical parameters providing the best fit of Eq. (\ref{fitepsn}) to the data obtained using the entropy theory).
\begin{figure}[ht]
\renewcommand{\figurename}{FIG.}
\begin{center}
\unitlength 1in
\includegraphics[width=2.8in]{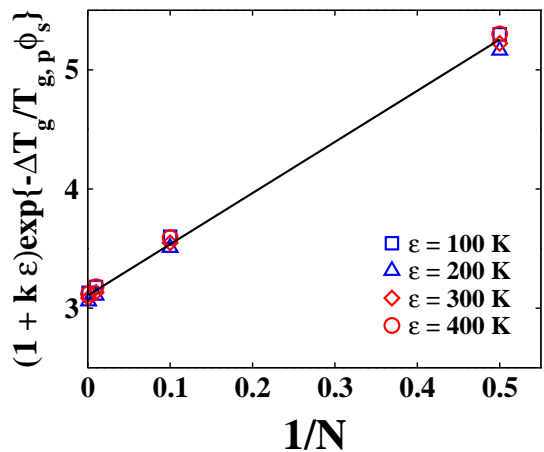}
\caption{\label{fig7} General scaling relation for the shift in the glass transition temperature $\Delta T_g$ for mixtures of a SF polymer melt diluted by an oligomeric additives when both the size $N$ and interaction energy $\epsilon$ of the diluent vary. The relation has the form $\Delta T_g = - T_{g, \, p} \phi_s \, \text{ln} \{(a + b/N)/(1 + k \, \epsilon)\}$ with $a = 3.11 \pm 0.04$, $b = 4.30 \pm 0.09$, and $k = (0.53 \pm 0.01)\times 10^{-2}$.}
\end{center}
\end{figure}
Equation (\ref{fitepsn}) implies that the exponentiated absolute value of the relative drop in $T_g$ scaled by the factor $f(\epsilon) = 1 + k \epsilon$ should be a linear function of inverse segment size $1/N$ of the additive, where $k$ is a properly chosen constant, and all computed data for different $N$ and $\epsilon$ from Fig. 2 can be scaled by an $\epsilon$-dependent linear factor $f(\epsilon)$ to fall on a single master curve which is a linear function of the inverse solvent size $1/N$.

Figure 8 depicts the dependence of the change in fragility $\Delta K = K - K_p$ of a mixture of the host semi-flexible SF polymer with 10 $\%$ mass of a fully flexible oligomeric species for different diluent sizes and cohesive energies plotted as a function of the change in excess free volume fraction at the respective $T_g$, $\Delta \phi_v = \phi_v (T_g) - \phi_{v, \, p}(T_{g, \, p})$.
\begin{figure}[ht]
\renewcommand{\figurename}{FIG.}
\begin{center}
\unitlength 1in
\includegraphics[width=3.0in]{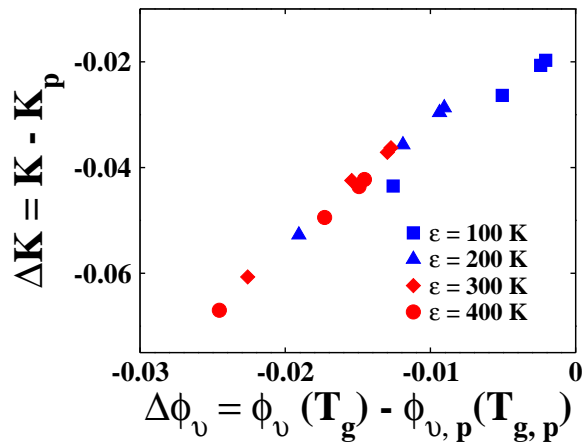}
\caption{\label{fig8} Correlation between the change in fragility $\Delta K = K - K_p$ of a SF polymer melt diluted by 10 $\%$ mass of an oligomeric species and the change in free volume fraction $\Delta \phi_v$ at the respective $T_g$.}
\end{center}
\end{figure}
Although the data for mixtures with different $\epsilon$ do not merge onto a single master curve, there is a strong correlation between the fragility $K$ and the excess free volume fraction $\phi_{v}(T_g)$, a trend  qualitatively analogous to the behavior found for the fragility $K_p$ of polymer $\it{melts}$, establishing a correlation of $K_p$ with $\phi_v(T_{g, \, p})$ as one of the parameters ($N$, $\epsilon$,  or $E_b$) varies for a given molecular topology. \cite{stukalin09} Thus, the general trends in the variation of $T_g$ and the fragility for glass formation in mixtures are qualitatively similar to the trends revealed for pure polymer melts. For instance, there is an evident analogy between (a) the influence of the size of diluents on $T_g$ and the fragility $K$ for the mixtures and on (a') the effect of the chain size on $T_{g, \, p}$ and $K_p$ for polymer melts. Similarly, changes in the cohesive energy affect $T_g$ and the fragility $K$ identically when (b) $\epsilon_{ss}$ of the diluent in binary fluids and (b') the cohesive energy $\epsilon_{pp}$ in the polymer melt increases (or decreases).

\section{PS-diluent mixtures: the efficiency of plasticization and antiplasticization} 

Small amounts of a plasticizer lead to a shift in $T_g$ of the binary fluid that is usually linear in the mass fraction $\phi_s$
of the additive, 
\begin{equation}\label{lintg}
T_g = T_{g, \, p} - \lambda_0 \, \phi_s,
\end{equation}
where $\lambda_0$ is  defined as the ``plasticizer efficiency parameter''. In particular, Mauritz \textit{et al.} develop this linear relation from diffusion theory and use it to model (poly)vinylchoride (PVC) plasticized by di-alkyl phthalates \cite{mauritz90a, mauritz90b}. However, Eq. (\ref{lintg}) fails at higher concentrations (above 40 $\%$ mass of diluent), where the dependence of $T_g$ on dilution generally becomes nonlinear \cite{beirnes86, martin03} (see Sect. VI).

A series of entropy theory computations for PS ($N = 8000$) diluted by 0 $\%$ to 20 $\%$ mass of an oligomeric species with different properties ($N$, $\epsilon$) reveals whether the general trends found for SF polymer systems diluted by a $\it{fixed}$ amount of additive display analogous behavior for the plasticizer (or antiplasticizer) efficiency parameters $\lambda_0$. PS is chosen to facilitate some comparison with experimental data because PS is one is the most extensively studied plasticized polymers. Due to limitations in modeling the monomer structure of PS (and of some diluents as well), the focus in the present work is on describing general trends rather than in quantitatively reproducing the experimental data. The plasticizers are considered as fully flexible oligomeric species, i.e., with the styrene subunit structure (Fig. 1). Figures 9 to 11 describe the general trends as follows: The composition dependence of $T_g$ for a mixture of PS and 0 $\%$ to 20 $\%$ mass of diluents is almost linear.
\begin{figure}[ht]
\renewcommand{\figurename}{FIG.}
\begin{center}
\unitlength 1in
\includegraphics[width=3.0in]{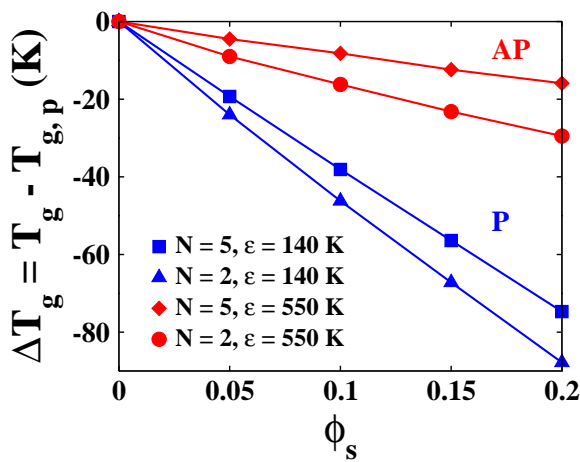}
\caption{\label{fig9} $T_g$ of polystyrene (PS) diluted by either a fully flexible pentamer ($N=5$) or a dimer ($N=2$) with the structure of PS but with low and high van der Waals energies $\epsilon$ as a function of the composition of the mixture.}
\end{center}
\end{figure}
\begin{figure}[h]
\renewcommand{\figurename}{FIG.}
\begin{center}
\unitlength 1in
\includegraphics[width=3.0in]{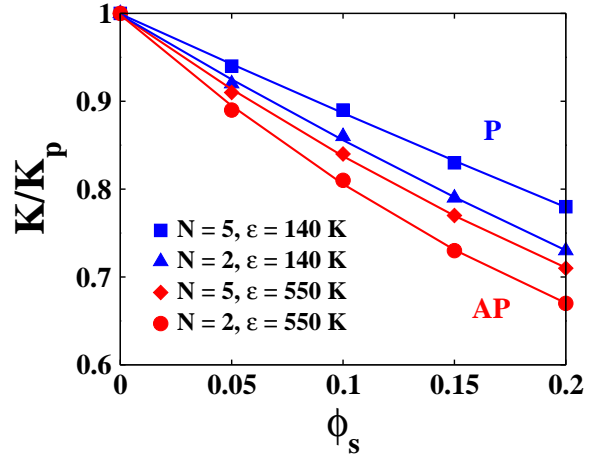}
\caption{\label{fig10} Composition dependence ($\phi_s$) of the fragility $K=1/D$ of a PS melt diluted by a fully flexible N=5 and N=2 oligomeric additive with low and high cohesive energies $\epsilon$.}
\end{center}
\end{figure}
\begin{figure}[ht]
\renewcommand{\figurename}{FIG.}
\begin{center}
\unitlength 1in
\includegraphics[width=3.0in]{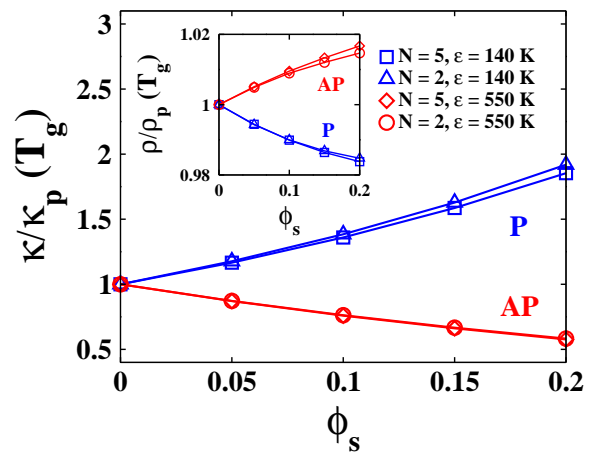}
\caption{\label{fig11} Composition dependence ($\phi_s$) of the ratios of isothermal compressibility of the mixtures (PS diluted by fully flexible N=5- and N=2-mers with low and high cohesive energies $\epsilon$) and the undiluted polymer, evaluated at the $T_g$ of the mixtures. The inset presents the identical relations for the specific density ratios at the $T_g$ of the mixtures.}
\end{center}
\end{figure}
In accord with results in Sect. III, the most efficient plasticizers (large $\lambda_0$) are those with the smallest $\epsilon$ and small sizes. Weak polymer-additive interactions ``soften'' the polymer matrix more than the undiluted material.  The ``compactness'' of the additive also is the primary feature driving the increase of $\lambda_0$ for antiplasticization. Strongly attractive polymer-diluent cross-interactions $\epsilon_{ps}$ lead to a toughening of the material but diminish the efficiency of the depression of $T_g$ (see Fig. 9, 11).  Thus, the rate of decrease of $T_g$ with dilution tends to be smaller for antiplasticized polymers, as demonstrated in Fig. 9. The computed $\lambda_0$ vary from 81 to 446, while experiments for PS indicate that $\lambda_0$ lies in the range from 241 to 524 depending on solvent/plasticizer \cite{martin03, jenckel53}. Both plasticizers and antiplasticizers reduce the fragility of the host polymer (see Fig. 10). The fragility parameter $m$ is also almost a linear function of $\phi_s$ for relatively low amounts of additives (within the diluent range from 0 $\%$ to 20 $\%$ mass),
\begin{equation}\label{fragmcomp} m = m_p - \eta_0 \, \phi_s,
\end{equation}
with the slopes $\eta_0$ varying from 134 to 202 in the model computations.

\section{Composition profile of configurational entropy in mixtures: $T_g$ of the plasticized polymer}

We have computed configurational entropies for fully miscible binary fluids comprising high molar mass semi-flexible polymers and low molar mass diluents over a wide range of compositions. Data of this type are available, for instance, for PS-plasticizer mixtures where the diluent is a solvent, such as toluene, chloroform, carbon disulfide, etc. The previous discussion focuses on the range from 0 $\%$ to 20 $\%$ mass of the additives where the composition dependence of $T_g$ emerges as linear to good accuracy (when the diluents are \textit{fully flexible} species).  To check whether the composition profile of $T_g$ levels off at high diluent content (as observed for many polymer-solvent mixtures \cite{jenckel53, pezzin68, beirnes86}), we have performed entropy theory computations for PS diluted by low molar mass ($N =5$) \textit{semi-flexible} ($E_b = 600$ K, $E_s = 200$ K)  oligomers (denoted as species A) to mimic binary fluids  in which the additive A has a (measurable) glass transition temperature ($T_{g, s} = 200$ K), which is typically much lower than the transition temperature of the host polymer ($T_{g, p} = 378$ K).  The monomers of A have the structure of 1-pentene and a cohesive energy ($\epsilon = 200$ K) appreciably smaller than the one for PS (see section II). Thus, A is expected to behave as a typical plasticizer.

The calculations demonstrate that the inverse of the ``exponential'' factor $1/\Omega (\phi_p) = (\beta \Delta \mu)^{-1} [s_c/s_c^{*}]$ in the Adam-Gibbs relation is to good accuracy a linear function of polymer mass fraction $\phi_p$ over the whole composition range at fixed $T$ (see Fig. 12).
\begin{figure}[ht]
\renewcommand{\figurename}{FIG.}
\begin{center}
\unitlength 1in
\includegraphics[width=3.0in]{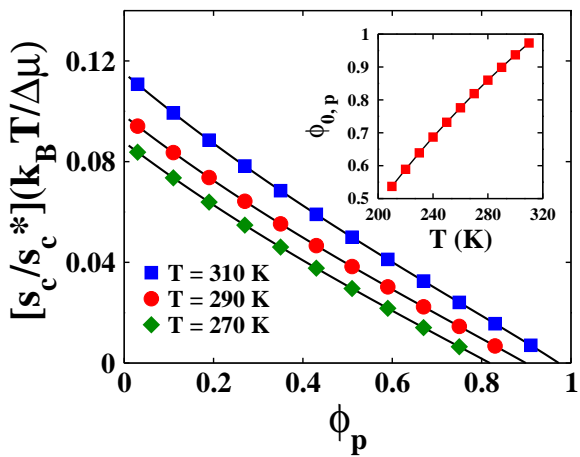}
\caption{\label{fig12} Dependence of the inverse ``exponential'' factor $1/\Omega (\phi_p) = (\beta \Delta \mu)^{-1} [s_c/s_c^{*}]$ in PS-A  mixtures (see text) at several  fixed temperatures as a function of the mass fraction $\phi_p$ of PS. The inset depicts the temperature dependence of the ``zero mobility concentration'' of the host polymer in PS-A mixtures.}
\end{center}
\end{figure}
This factor $\Omega$ combines the concentration dependence of the configurational entropy $s_c$ of a binary fluid, its high temperature limiting value $s_c^{*}$, and the high temperature activation energy for the PS-A mixtures $\Delta \mu \simeq 6 T_I$, which is computed using the same empirically based relation as for the melts. The crossover temperatures $T_I$ as well as the limiting value $s_c^{*}$ are found to be linear functions of the composition of the binary mixtures to good accuracy. Therefore, our ETGF computations suggest that the high temperature activation energy for binary fluids $\Delta \mu$ varies linearly with additive concentration
\begin{equation}\label{deltamucomp}
\Delta \mu (\phi_s) = (1 - \phi_s) \Delta \mu_p + \phi_s \Delta \mu_s,
\end{equation}
For instance, for PS-A mixtures the computed $\Delta \mu$ varies from 25.6 kJ/mol for PS to 16.5 kJ/mol for A. 
The configurational entropy of a binary mixture scales linearly with composition, $\phi_p$ as $s_c \simeq \alpha (\phi_{0, \, p} - \phi_p)$. Our computations show that the overall composition dependence of the inverse ``exponential'' factor $1/\Omega(\phi_p)$ at fixed temperature is also linear with slight deviations from linearity  at high dilutions (see Fig. 12). The linearity of $1/ \Omega$ in composition $\phi_p$ implies that the computed structural relaxation time $\tau_{\alpha}(\phi_p) = \tau_0 \, \text{exp} \{ \Omega (\phi_p) \}$ exhibits a composition dependence at fixed temperature that accurately follows the Angell equation, \cite{gordon77, miura92, adachi75}
\begin{eqnarray}\label{angell}
\tau_{\alpha}(\phi_p)& = & \tau_0 \, \text{exp} \{ \beta \Delta \mu(\phi_p) [s_c^{*}(\phi_p)/s_c(\phi_p)] \} \nonumber \\
& = & A \, \text{exp} \{B/(\phi_{0, \, p} - \phi_p) \},
\end{eqnarray}
where $\phi_p$ is the polymer concentration in the mixture, $\phi_{0, \, p}$ is its ``critical'' concentration at which the extrapolated  relaxation time diverges (the extrapolated \textit{zero mobility} concentration), and $A$ and  $B$ are some concentration independent constants.

An analysis of the computed $\tau_{\alpha} (\phi_p)$ over the whole composition range yields the variation  of $T_g$ for PS-A mixtures in Fig. 13, where the full composition profile of $T_g$ is generally nonlinear.
\begin{figure}[ht]
\renewcommand{\figurename}{FIG.}
\begin{center}
\unitlength 1in
\includegraphics[width=3.0in]{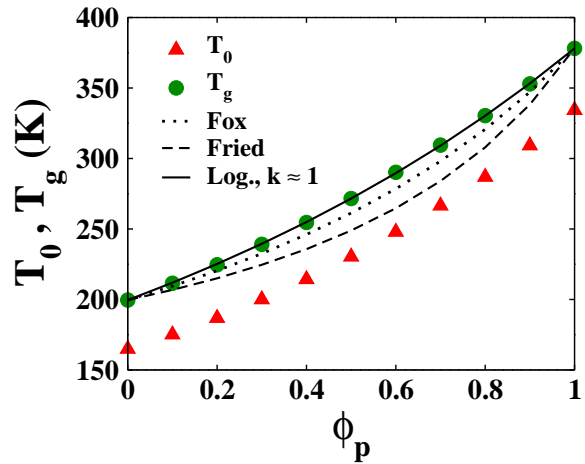}
\caption{\label{fig13} Compositional variation of the glass transition temperature $T_g$ for PS-A mixtures, where $\phi_p$ is the mass fraction of the polymer.}
\end{center}
\end{figure}
The composition dependent efficiency parameter $\lambda (\phi_s) = - d T_g / d \phi_s$ of the plasticizer A progressively diminishes as the amount of diluent increases, a behavior that accords with the observed compositional variation of $T_g$ for polystyrene (PS) in 12 different solvents, where a continuously decreasing negative slope $d T_g / d \phi_s$ is indicated. \cite{plazek96} However, in experiments for small amounts of solvent mixed with PS and in our computations for PS diluted by 0 $\%$ to 20 $\%$ mass of A, the slope $\lambda (\phi_s)$ is constant, so the depression in $T_g$ for the PS-A mixtures may be evaluated as $\Delta T_g \simeq - \lambda_0 \, \phi_s$. \cite{martin03} Jenckel and Heush have suggested the expression, \cite{jenckel53, kalogeras09}
\begin{equation}\label{jenckel}
T_g = (1 - \phi_s) T_{g, \, p} + \phi_s T_{g, \, s} + b (T_{g, \, p} - T_{g, \, s}) \phi_s (1 - \phi_s),
\end{equation} 
for plasticized polymer fluids with $b$ a parameter that characterizes the solvent quality of the plasticizer. A frequently used relation that enables prediction of binary fluid properties from the properties of pure components is the Fox equation \cite{fox56, kalogeras09} that is derived by assuming random mixing between the components, equal heat capacity jumps in the glass transition region, and vanishing excess volume of mixing over the entire concentration range, to yield
\begin{equation}\label{fox}
\frac{1}{T_g} = \frac{1 - \phi_s}{T_{g, \, p}} + \frac{\phi_s}{T_{g, \, s}}.
\end{equation}
Among the other proposed expressions for $T_g$ of mixtures that also do not require adjustable parameters is the one suggested by Fried, \cite{fried82}
\begin{equation}\label{fried}
\text{ln}(T_g/T_{g, \, p}) = \frac{\phi_s \text{ln}(T_{g, \, s}/T_{g, \, p})}{(1 - \phi_s) (T_{g, \, s}/T_{g, \, p}) + \phi_s},
\end{equation}
which likewise may be used to evaluate $T_g$ of the mixture if $T_{g, \, p}$ and $T_{g, \, s}$ of the pure components  are known. A variant of this equation \cite{ceccorulli87} contains an extra adjustable parameter $k_1$ replacing $T_{g, \, s}/T_{g, \, p}$ in the denominator of Eq. (\ref{fried}) to get
\begin{equation}\label{logexp}
\text{ln}(T_g/T_{g, \, p}) = \frac{\phi_s \text{ln} (T_{g, \, s}/T_{g, \, p})}{(1 - \phi_s) k_1 + \phi_s}.
\end{equation}
We use the computed compositional dependence of $T_g$ for PS-A mixtures and the $T_g$ of the pure components to test Eqs. (\ref{fox}), (\ref{fried}), and (\ref{logexp}). Both equations (\ref{fox}) and (\ref{fried}) with no adjustable parameters predict a diminishing of plasticizer efficiency $\lambda$ as the amount of diluent in the mixture increases. The Fried equation does not describe our computations well, while the $T_g$ prediction by the Fox equation closely corresponds to the predictions of the entropy theory (see Fig. 13). Our data can be fit excellently by the logarithmic relation in Eq. (\ref{logexp}) with $k_1 = 0.93$ (in contrast, $T_{g, \, s}/T_{g, \, p} \simeq 0.53$). Analogously, the Vogel-Fulcher-Tammann temperature $T_0$ of the PS-A mixtures can be described by the relation, 
\begin{equation}\label{logexp2}
\text{ln}(T_0/T_{0, \, p}) = \frac{\phi_s \text{ln} (T_{0, \, s}/T_{0, \, p})}{(1 - \phi_s) k_1 + \phi_s},
\end{equation}
that is identical to Eq. (\ref{logexp}) except for the use of a slightly different $k_1 = 0.89$ from the fit (not shown in Fig. 13). Since the adjustable parameter $k_1$ is close to unity $k_1 \simeq 1$, setting $k_1 = 1$ in Eq. (\ref{logexp}) produces the simpler relation for compositional dependence of $T_g$ for the mixtures as, \begin{equation}\label{logsimp}
T_g =  T_{g, \, p}^{1 - \phi_s} \, T_{g, \, s}^{\phi_s},
\end{equation} 
and the plasticizer efficiency parameter is
\begin{equation}\label{logsimp2}\lambda (\phi_s) = - d T_g/d \phi_s =  T_{g, \, p}^{1 - \phi_s} \, T_{g, \, s}^{\phi_s} \, \text{ln} (T_{g, \, p}/T_{g, \, s}). 
\end{equation} 
When the diluent concentration is small, $\lambda (\phi_s) \simeq \lambda (0) \equiv \lambda_0$, and the glass transition depression $\Delta T_g = T_g - T_{g, p} \simeq - \lambda_0 \phi_s$ can be evaluated using the simpler relation,
\begin{equation}\label{great}
\Delta T_g = - \phi_s \, T_{g, \, p} \, \text{ln} (T_{g, \, p}/T_{g, \, s}),
\end{equation} 
which reproduces computations of $\Delta T_g$ for PS diluted by 0 $\%$ to 30 $\%$ mass of A with relative errors not exceeding 5.6 $\%$. The plasticizer efficiency parameter $\lambda$ derived for A can be generalized because the decrease in $T_g$ upon dilution occurs for both plasticization and antiplasticization. Hence, the influence of diluent properties on the $T_g$ of mixtures may be understood by inserting additional relations into Eq. (\ref{great}) describing the dependence of the pure component $T_{g, \, s}$ and $T_{g, \, p}$ on various properties using relations that have been derived/observed when the diluents are \textit{oligomeric} species. For instance, the chain length dependence of $T_g$ from experimental measurements and from our computations is given by the relation $1/T_{g} = 1/T_{g}^{\infty} + K/N$. Additionally, for a given molecular topology, $T_g$ scales approximately linearly with the strength of the van der Waals energy $\epsilon$ as $T_g = A + k \, \epsilon$ (see, for instance, Fig. 3 in Ref. 3).  Combining the general $N$- and $\epsilon$-scaling relations for $T_g$ with Eq.  (\ref{great}) yields Eq. (\ref{fitepsn}), which rationalizes the influence of diluent properties on $T_g$ for binary mixtures and successfully reproduces our model computations for a SF polymer diluted by a fixed (small) amount of oligomeric species (see section II). Although, these diluents are treated as fully flexible and thus have no definite glass transition temperature, the general trends are excellently described by Eq. (\ref{great}) that has been ``derived'' from an analysis of the compositional variation of $T_g$ for PS-A mixtures.

\section{Zero mobility concentration: dependence on temperature, pressure and chain length}

The ``zero mobility concentration'' $\phi_{0, \, p}$ generally depends on temperature in a fashion that can be deduced from the entropy theory computations. The inset of Fig. 12 depicts the ``zero mobility concentration'' $\phi_{0, \, p}$  for PS-A mixtures at ambient pressure ($P \simeq 0$) as a function of temperature for $T < T_{0, \,  p}$, where $T_{0, \, p}$ is the VFT temperature \cite{ferry80} of the undiluted PS polymer melt ($T_{0, \, p} \simeq 330$ K). The region $T > T_{0, \, p}$ is omitted because  it lacks an extrapolated divergence of the structural relaxation times (no entropy catastrophe) for any composition of the PS-A mixture. Thus, $\phi_{0, \, p}$ in the Angell equation cannot strictly be interpreted as a ``zero mobility concentration'' for $T > T_{0, \, p}$, as in the case of the VFT temperature; $T_{0, \,  p}$ only has meaning in the sense of extrapolation. $\phi_{0, \, p}$ diminishes monotonically as temperature decreases, in accord with expectations that if the glass formation occurs at \textit{fixed} temperature by varying the composition of binary fluid (e.g., by gradual evaporation of the solvent) than less diluent $\phi_s = 1 - \phi_p$ is left in the mixture in state $X_0$ when the (extrapolated) divergence of $\tau_{\alpha}(\phi_p)$ is reached at higher  temperature and more solvent remains in the fluid in this state $X_0$ at lower temperature.

Our ETGF theory allows determining the pressure dependence of the ``zero mobility concentration''. These computations are illustrated for PS-A mixtures in the temperature range $T$ from 250 K to 310 K and a wide pressure range $P$ from 0 MPa to 500 MPa. Basically, $\phi_{0, \, p}$ is linear at low pressures and tend to saturate at high pressures.
\begin{figure}[ht]
\renewcommand{\figurename}{FIG.}
\begin{center}
\unitlength 1in
\includegraphics[width=3.0in]{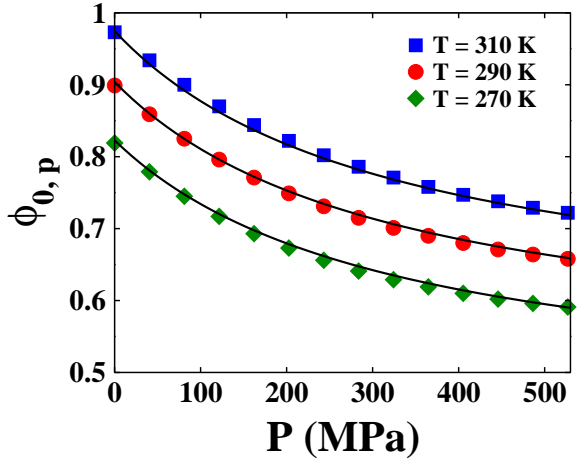}
\caption{\label{fig14} Pressure dependence of ``zero mobility concentrations'' $\phi_{0, \, p}$ of the polymer in PS-A mixtures at several fixed temperatures. Solid lines represent the non-linear fits to Eq. (\ref{xpt}) in the range from 0 MPa to 500 MPa and from 250 K to 310 K with parameters: $\phi_0 = 1$, $T_{0} = 317.9$ K, $\phi_{\infty} = 1.335$, $T_{\infty} = 239.6$ K, $h = 0.003$ MPa$^{-1}$.}
\end{center}
\end{figure}
The initial slope ${\partial \phi_{0, \, p}}/{\partial P}$ for PS-A mixtures is found to be ${\partial \phi_{0, \, p}}/{\partial P} \simeq  - 0.001$ MPa$^{-1}$ in the range of temperatures from 250 K to 310 K. The temperature coefficient ${\partial \phi_{0, \, p}}/{\partial T} \simeq 0.004$ K$^{-1}$ from the inset of Fig. 12, which only slightly varies for different pressures, combined with the low pressure coefficient ${\partial \phi_{0, \, p}}/{\partial P}$ from Fig. 14, enables us to evaluate the pressure dependence of the Kauzmann temperature $T_K$ at low pressures, i.e., the average pressure coefficient of the temperature at which the configurational entropy extrapolates to zero,  as ${\partial T_K}/{\partial P} = - ({\partial \phi_{0, \, p}}/{\partial P})/({\partial \phi_{0, \, p}}/{\partial T}) \simeq 0.25$ K MPa$^{-1}$, which is consistent with our direct computation of $dT_{K, \, p}/dP \simeq 0.30$ K MPa$^{-1}$ for \textit{pure} PS at low pressures. Moreover, this value is close to the reported experimental pressure dependence of the Vogel temperature $dT_{0, \, p}/dP \simeq 0.32 \pm 0.05$ K MPa$^{-1}$ for PS in the range from 10 MPa to 80 MPa \cite{sahnoune96}. The combined $T-P$ dependence of $\phi_{0,\, p}$ for PS-A mixtures (see Fig. 14) can be fit well by the expression
\begin{equation}\label{xpt}
\phi_{0,\, p} = \frac{1 + (\phi_{0} - T_{0}/T) + h(\phi_{\infty} - T_{\infty}/T)P}{1 + h \,P},
\end{equation} where
$\phi_{0}$,  $T_{0}$; and $\phi_{\infty}$, $T_{\infty}$; and $h$ are adjustable parameters. The first two parameters are found to equal $\phi_{0} \simeq \phi_{0, \, p, \, p} \equiv 1$ and $T_{0} \simeq T_{K, \, p}$ (see also the caption to Fig. 14 for numerical values). These parameters define the temperature dependence of $\phi_{0, \, p}$ at ambient pressure $P \simeq 0$, while the second pair of parameters are related to the behavior of $\phi_{0,\, p}$ in the high pressure limit. Lastly, the parameter $h \ll 1$ in Eq. (\ref{xpt}) is associated with the low pressure coefficient of the ``zero mobility concentration'', namely ${\partial \phi_{0, \, p}}/{\partial P} \simeq  - \kappa \, h$ with $\kappa = 1 + (\phi_{0} - \phi_{\infty}) + (T_{\infty} - T_0)/T$.  

The ``zero mobility concentration'' $\phi_{0, \, p}$ depends on the chain length of the host polymer. The computed $\phi_{0, \, p}$ at fixed temperature for different polymerization indices $N$ of the host polymer in PS-A mixtures scales as $\phi_{0, \, p} \sim 1/N$, analogous to the $T_{g, p} \sim 1/N$ scaling for undiluted polymer melts, giving 
\begin{equation}
\phi_{0, \, p} = \phi_{0, \, p}^{\infty} + \theta/N.
\end{equation} 
The slope $\theta$ of this dependence only changes slightly with temperature, as displayed in Fig. 15 where the extrapolated ``zero mobility concentration'' of PS is presented as a function of the inverse chain length $1/N$ of the host polymer for three different temperatures, $T < T_{0, \, p}$.
\begin{figure}[ht]
\renewcommand{\figurename}{FIG.}
\begin{center}
\unitlength 1in
\includegraphics[width=3.0in]{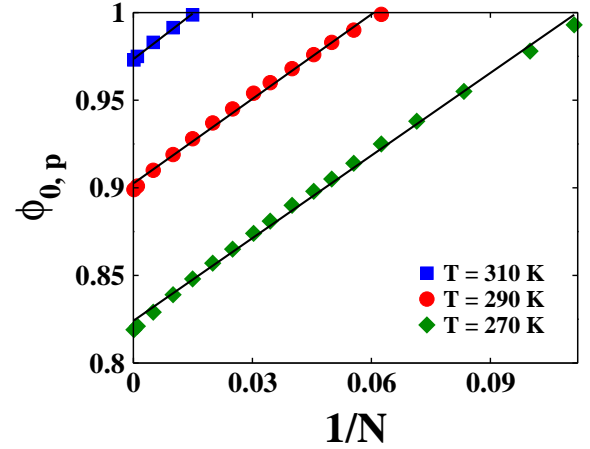}
\caption{\label{fig15} Chain length dependence of the ``zero mobility concentrations'' $\phi_{0, \, p}$ of the polymer in PS-A mixtures at several fixed temperatures. Solid lines represent the linear regressions $\phi_{0, \, p} = \phi_{0, \, p}^{\infty} + \theta/N$ with parameters $\phi_{0, \, p}^{\infty} = 0.973$ and $\theta = 1.739$ ($T = 310$ K); $\phi_{0, \, p}^{\infty} = 0.902$ and $\theta = 1.608$ ($T = 290$ K); $\phi_{0, \, p}^{\infty} = 0.824$ and $\theta = 1.572$ ($T = 270$ K)}
\end{center}
\end{figure}
Equation (\ref{great}) predicts that the drop in $T_g$ upon dilution is proportional to the glass transition temperature $T_{g, \, p}$ of the host polymer. Hence, the plasticization efficiency $\lambda (N_l)$ of a given plasticizer for a lower molar mass host polymer should be smaller than $\lambda (N_h)$ for a host polymer with higher molar mass by a factor equal to the ratio of their glass transition temperatures $T_{g, \, p}$. To confirm this prediction, we have computed $\Delta T_g$ of a low molar mass PS ($N_l = 20$; $T_{g, \, p} = 361.3 $ K) diluted by 20 $\%$ mass of the diluent A which is an oligomer of the high molar mass PS ($N_h = 8000$; $T_{g, \, p} = 378.1$ K).  The computed shift $\Delta T_g = - 44.1$ K for the low molar mass PS is close to the $\Delta T_g = - 45.6$ K predicted using the $T_{g, \, p}$-scaling in the relation Eq. (\ref{great}) and the computation for the high molar mass PS ($\Delta T_g = - 47.7$ K). When the molar mass of PS drops to $N_l$, the predicted shift in $T_g$ for the PS-A mixture is small because our entropy theory underestimates the experimentally observed slope of the $1/T_{g, \, p}$ vs. $1/N$ relation for PS systems by a factor of 3. Using the experimental dependence of $T_{g, \, p}$ on molar mass for PS leads to the expectation of a bit larger change in the plasticizer efficiency parameter $\Delta \lambda / \lambda \simeq 0.1$ with variation of the molar mass from very high polymerization indices down to $N_l = 20$ for this example. However, the generally favorable qualitative comparison between the observed and calculated $\Delta T_g$ for mixtures encourages the belief that Eq. (\ref{great}) is universal enough to capture the general trends for glass formation in binary mixtures, not only when the properties of diluents vary (as shown in section II), but also when \textit{some} properties the host polymer change (e.g., the chain length).

\section{Conclusions}

Systematic computations with the generalized entropy theory provide significant insight into the influence of additives on glass formation in polymers. Our studies reveal general guidelines concerning the molecular basis for both plasticization and antiplasticization. The computations establish the primarily role of diluent's size and its interaction energy with the host polymer in creating conditions that allow either plasticization or antiplasticization of the polymeric material. Mixing flexible diluent molecules having small size and strong attractive interactions with the polymer matrix converts the fragile polymer system into a stronger glass-former with antiplasticization characteristics, a finding in accord with recent simulations of Riggleman \textit{et al.} \cite{riggleman06} In contrast, the addition small, flexible diluents having weak attractive interactions with a host polymer leads to plasticization and a reduction in fragility. The computed efficiencies of the plasticizer or antiplasticizer are rationalized in terms of simple scaling relations that describe the influence of solvent size and interactions on the variation of the glass transition temperatures with diluent concentration. Changes in the isothermal compressibility and specific density upon dilution tend to be uniform at all temperatures. For instance, antiplasticized mixtures densify on mixing and are less compressible at all temperatures than the undiluted polymer melt. However, computations of the isothermal compressibility and density rather than the bulk modulus cannot reveal the existence of an ``antiplasticization'' temperature $T_a$, below which antiplasticization is observed and above which only plasticization appears. \cite{anopchenko06}

A simple scaling relation is found to describe the compositional variation of the computed glass transition temperatures of plasticized polymers in which $T_g$ for the mixtures is reconstructed from the $T_g$ of the pure components. This scaling relation is consistent with experimental observations in predicting that the (anti-)plasticizer efficiency parameters continuously diminish as the fluid becomes more diluted. An analogous relation is deduced for the Vogel-Fulcher-Tammann temperatures $T_0$ as a function of composition. The structural relaxation times are predicted to obey the Angell equation found previously by analyzing experimental data for the viscosity and electrical conductivity of binary mixtures in the supercooled region.\cite{angell72}  The ``zero mobility concentration'' $\phi_{0, \, p}$, at which the extrapolated relaxation time diverges, depends approximately linearly on temperature and on the inverse chain length of the host polymer. 

Further research using the entropy theory for mixtures should be aimed at resolving the issue concerning whether two glass transition temperatures occur when the fluid mixture exhibits phase separation. \cite{dudowicz93}  In this case, the criterion $\tau_{\alpha} \sim O$(100 s) can be applied to determine both temperatures simultaneously. (However, the criterion cannot be applied if the fluid has two transitions in the one-phase region). Our computations for fully miscible PS-A mixtures do not suggest a sharp variation of the configurational entropy of mixtures for temperatures comparable to the $T_g$ of each pure components, even when the pure system $T_g$ differ by more than 170 K. 

Another interesting aspect for study is the influence of non-Berthelot values of the exchange energy $\epsilon = \epsilon_{pp} + \epsilon_{ss} - 2 \, \epsilon_{ps}$. We have invoked the Berthelot constraint to restrict the number of interaction energy parameters, and lifting this condition should have interesting consequences and provide the influence of the exchange energy on glass formation in binary fluids for a given architecture of the solvent molecule and for properties of a host polymer.  \\

This research is supported, in part, by NSF grant CHE-0749788, PRF grant 46666-AC7, and by the Joint Theory Institute which is funded by the University of Chicago and Argonne National Laboratory. We are grateful to Jacek Dudowicz for several helpful discussions. 

\section*{}

\end{spacing}

\end{document}